\documentclass[a4paper,12pt]{article}
\usepackage[pctex32]{graphics}

\begin{document}
\topmargin 0pt \oddsidemargin 0mm

\renewcommand{\thefootnote}{\fnsymbol{footnote}}
\begin{titlepage}
\begin{flushright}
INJE-TP-06-08\\
gr-qc/0612112
\end{flushright}

\vspace{5mm}
\begin{center}
{\Large \bf Non-flat universe and interacting  dark energy model }
\vspace{12mm}

{\large  Kyong Hee Kim, Hyung Won Lee  and Yun Soo
Myung\footnote{e-mail
 address: ysmyung@inje.ac.kr}}
 \\
\vspace{10mm} {\em  Institute of Mathematical Sciences and School
of Computer Aided Science \\ Inje University, Gimhae 621-749,
Korea}

\end{center}

\vspace{5mm} \centerline{{\bf{Abstract}}}
 \vspace{5mm}
For non-flat universe of $k\not=0$, we investigate  a model of the
interacting holographic dark energy with cold dark matter (CDM).
There exists a mixture of two components arisen from decaying of
the holographic dark  energy into CDM. In this case we use the
effective equations of state ($\omega^{\rm eff}_{\rm
\Lambda},~\omega^{\rm eff}_{\rm m}$) instead of the native
equations of state ($\omega_{\rm \Lambda},\omega_{\rm m})$.
 Consequently, we show that
interacting holographic energy models in non-flat universe cannot
accommodate a transition from the dark energy to the phantom
regime.
\end{titlepage}
\newpage
\renewcommand{\thefootnote}{\arabic{footnote}}
\setcounter{footnote}{0} \setcounter{page}{2}

\section{Introduction}
Recent observations from Supernova (SN Ia)~\cite{SN} and large
scale structure~\cite{SDSS} imply that our universe is
accelerating.   Also cosmic microwave background
observations~\cite{Wmap1,Wmap3} provide an evidence for the
present acceleration. A combined analysis of cosmological
observations shows that the present universe consists of 70\% dark
energy and 30\% dust matter including CDM and baryons.

Although there exist a number of dark energy models, a promising
candidate is the cosmological constant. However, one has the two
famous cosmological constant problems: the fine-tuning and
coincidence problems. In order to solve these problems, we need a
dynamical cosmological constant model derived  by the holographic
principle.  The authors in~\cite{CKN} showed that in quantum field
theory, the UV cutoff $\Lambda$ could be related to the IR cutoff
$L_{\rm \Lambda}$ due to the limit set by introducing  a black
hole (the effects of gravity). In other words, if $\rho_{\rm
\Lambda}=\Lambda^4$ is the vacuum energy density caused by the UV
cutoff, the total energy of system with the size $L_{\rm \Lambda}$
should not exceed the mass of the black hole with the same size
$L_{\rm \Lambda}$: $L_{\rm \Lambda}^3 \rho_{\rm \Lambda}\le
2M_p^2L_{\rm \Lambda}$. If the largest cutoff $L_{\rm \Lambda}$ is
chosen to be the one saturating this inequality,  the holographic
energy density is  given by $\rho_{\rm \Lambda}= 3c^2M_p^2/8\pi
L_{\rm \Lambda}^2$ with a constant $c \ge 1$. The lower limit of
$c$ is protected by the entropy bound. Here we regard $\rho_{\rm
\Lambda}$ as a dynamical cosmological constant. Taking $L_{\rm
\Lambda}$ as the size of the present universe, the resulting
energy  is close to the present dark energy~\cite{HMT}. However,
this
 approach with $L_{\rm
\Lambda}=1/H$ is not complete  because it fails to recover the
equation of state (EoS) for the dark energy-dominated
universe~\cite{HSU}. Further studies in \cite{LI,FEH,Myung2,Zim3}
have shown that choosing the future event horizon as the IR cutoff
leads to an accelerating universe with $\omega_{\rm
\Lambda}=-1/3-2\sqrt{\Omega_{\rm \Lambda}}/3c$.

On the other hand, the interacting dark energy models provided a
new direction to understand the dark energy~\cite{Hor,Szy,Cop}.
The authors in \cite{WGA} introduced an interacting holographic
dark energy model where an interaction exists between holographic
energy  and CDM. They derived the phantom-phase of $\omega_{\rm
\Lambda}<-1$ using the native EoS $\omega_{\rm \Lambda}$. However,
it turned out that
 the interacting holographic dark energy model could not  describe a
phantom regime of $\omega^{\rm eff}_{\rm \Lambda}<-1$ when using
the effective equation of state  $\omega^{\rm eff}_{\rm
\Lambda}$~\cite{KLM}.  A key of this system is an interaction
between two matters. Their contents are changing due to energy
transfer from holographic energy to CDM until the two components
are comparable. If there exists a source/sink in the right-hand
side of the continuity equation, we must be careful to define its
EoS. In this case the effective EoS is the only candidate to
represent the state of the mixture of two components arisen from
decaying of the holographic energy into CDM. This is
 different  from the non-interacting case which is described by the native EoS.
More recently, it was shown  that for non-flat universe of
$k\not=0$~\cite{HL,WLA}, the interacting holographic dark energy
model could not describe a phantom regime of $\omega^{\rm
eff}_{\rm \Lambda}<-1$~\cite{Seta}.

In this work, we wish to address this issue again because the
previous works contain a few of ambiguous points. We solve two
coupled differential equations for density parameters $\Omega_{\rm
\Lambda}$ and $\Omega_{\rm k}$ numerically. Furthermore, we
introduce a general form of interaction $Q$ to find the
CDM-dominated universe with $\omega^{\rm eff}_{\rm m}=0$ at the
far past. We confirm that the phantom-phase is not found from
interacting holographic dark energy models.

\section{Interacting model in non-flat universe}
 Let us imagine a universe made of CDM $\rho_{\rm m}$ with
$\omega_{\rm m}=0$, but obeying the holographic principle. In
addition, we propose that the holographic energy density
$\rho_{\rm \Lambda}$ exists with $\omega_{\rm \Lambda}\ge-1$. If
one assumes a form of the interaction $Q=\Gamma \rho_{\rm
\Lambda}$, their continuity equations take the forms
\begin{eqnarray}
\label{2eq1}&& \dot{\rho}_{\rm \Lambda}+3H(1+\omega_{\rm \Lambda})\rho_{\rm \Lambda} =-Q, \\
\label{2eq2}&& \dot{\rho}_{\rm m}+3H\rho_{\rm m}=Q.
\end{eqnarray}
 This shows that the mutual interaction could
provide a mechanism to the particle production. Actually, this is
a decaying of the holographic energy component into CDM with the
decay rate $\Gamma$.  Taking a ratio of two energy densities as
$r_{\rm m}=\rho_{\rm m}/\rho_{\rm \Lambda}$, the above equations
lead to
\begin{equation}
\label{2eq3} \dot{r}_{\rm m}=3Hr_{\rm m}\Big[\omega_{\rm \Lambda}+
\frac{1+r_{\rm m}}{r_{\rm m}}\frac{\Gamma}{3H}\Big]
\end{equation} which means that the evolution of the ratio depends on the explicit form of interaction.  Even
if one starts with $\omega_{\rm m}=0$ and $\omega_{\rm
\Lambda}=-1$, this process is necessarily accompanied by the
different equations
 of state  $\omega^{\rm eff}_{\rm m}$ and $\omega^{\rm eff}_{\rm \Lambda}$. The decaying process impacts
their equations of state  and particularly, it induces the
negative effective EoS of CDM.  Interestingly, an  accelerating
phase could arise from a large effective non-equilibrium pressure
 $\Pi_{\rm m}$  defined as $\Pi_{\rm m}\equiv -\Gamma\rho_{\rm \Lambda}/3H(=-\Pi_{\rm
 \Lambda})$. Then  the two  equations (\ref{2eq1})
and (\ref{2eq2}) are translated into those of the two
dissipatively imperfect fluids
\begin{eqnarray}
\label{2eq4}&& \dot{\rho}_{\rm \Lambda}+ 3H\Big[1+\omega_{\rm
\Lambda}+\frac{\Gamma}{3H} \Big]\rho_{\rm \Lambda}=\dot{\rho}_{\rm
\Lambda}+ 3H\Big[(1+\omega_{\rm \Lambda})\rho_{\rm
\Lambda}+\Pi_{\rm
 \Lambda}\Big]=0, \\
\label{2eq5}&& \dot{\rho}_{\rm m}+3H\Big[1-\frac{1}{r_{\rm
m}}\frac{\Gamma}{3H}\Big]\rho_{\rm m}=\dot{\rho}_{\rm
m}+3H(\rho_{\rm m}+\Pi_{\rm m})=0.
\end{eqnarray}
The positivity of $\Pi_{\rm \Lambda}>0$ shows a decaying  of
holographic energy density via the cosmic frictional force, while
$\Pi_{\rm m}<0$ induces a production of the mixture via the cosmic
anti-frictional force simultaneously~\cite{Zim1,myung}. This is a
sort  of the vacuum decay process to generate a particle
production within the two-fluid model~\cite{Zim2}. As a result, a
mixture of two components will be  created. From Eqs.(\ref{2eq4})
and (\ref{2eq5}), turning on the interaction term, we define their
effective equations of state as
\begin{equation}
\label{2eq6} \omega^{\rm eff}_{\rm \Lambda}=\omega_{\rm
\Lambda}+\frac{\Gamma}{3H},~~ \omega^{\rm eff}_{\rm
m}=-\frac{1}{r_{\rm m}}\frac{\Gamma}{3H}. \end{equation} In this
work, we choose the general decay rate of $\Gamma=3b^2(1+r_{\rm
m}){^n}H$ with the coupling constant $b^2$ and $n \le
1$~\cite{BS}. For $n>1$, $\omega^{\rm eff}_{\rm \Lambda}$ diverges
for small $\Omega_{\rm \Lambda}$, while for $n<1$, one finds
$\omega^{\rm eff}_{\rm m}=0$ for  $\Omega_{\rm \Lambda}=0$ which
is better in agreement with the data.  On the other hand, the
first Friedmann equation  for $k\not=0$ is given by
\begin{equation}
\label{2eq7} H^2=\frac{8\pi}{3M^2_p}\Big[
 \rho_{\rm \Lambda}+\rho_{\rm m}\Big]-\frac{k}{a^2}.
\end{equation}
 Differentiating
Eq.(\ref{2eq7}) with respect to the cosmic time $t$ and then using
Eqs.(\ref{2eq1}) and (\ref{2eq2}), one finds the second Friedmann
equation as
\begin{equation}
\label{2eq8} \dot{H}=- \frac{3}{2}H^2\Big[1+\frac{
 \omega_{\rm \Lambda}}{1+r_{\rm m}}\Big]- \frac{1}{2}\frac{k}{a^2}\Big[1+\frac{
 3\omega_{\rm \Lambda}}{1+r_{\rm m}}\Big]
\end{equation}
which is useful to study the evolution when choosing $L_{\rm
\Lambda}=1/H$.
 Let us
introduce density parameters
\begin{equation}
\label{2eq9} ~\Omega_{\rm m}=\frac{8\pi \rho_{\rm
m}}{3M_p^2H^2},~\Omega_{\rm \Lambda}=\frac{8 \pi \rho_{\rm
\Lambda}}{3M^2_pH^2},~ \Omega_{\rm
k}=\frac{k}{a^2H^2}\end{equation} which allow to rewrite the first
Friedmann equation as
\begin{equation} \label{2eq10} \Omega_{\rm m}+\Omega_{\rm
\Lambda}=1+\Omega_{\rm k}.\end{equation}
 Then we can express $r_{\rm m}$ and $r_{\rm k}=\rho_{\rm
k}/\rho_{\rm \Lambda}$ in terms of $\Omega_{\rm \Lambda}$ and
$\Omega_{\rm k}$ as
\begin{equation}
\label{2eq11} r_{\rm m}=\frac{1-\Omega_{\rm \Lambda}+\Omega_{\rm
k}}{\Omega_{\rm \Lambda}},~~r_{\rm k}=\frac{\Omega_{\rm
k}}{\Omega_{\rm \Lambda}}.
\end{equation}

\section{Non-flat universe with the future event horizon}

In the case of $\rho_{\rm \Lambda}$ with Hubble horizon ($L_{\rm
\Lambda}=1/H$), we always have a fixed ratio $r_{\rm m}$ of two
energy densities. This provides the same negative EoS for both two
components~\cite{Zim3,myung}. For a null geodesic, we  introduce
the future event horizon $L_{\rm \Lambda}=R_{\rm FH}=a\chi_{\rm
FH}(t)=a\chi^{k}_{\rm FH}(t)$ with~\cite{Wein}
\begin{equation}
\label{3eq1} \chi_{\rm FH}(t)=\int_t^{\infty} \frac{dt}{a}.
\end{equation}
Here the comoving horizon size is given by
\begin{equation} \label{3eq2}
\chi^k_{\rm
FH}(t)=\int_{0}^{r(t)}\frac{dr}{\sqrt{1-kr^2}}=\frac{1}{\sqrt{|k|}}{\rm
sinn}^{-1}\Bigg[\sqrt{|k|}r(t)\Bigg],
\end{equation}
where leads to $\chi^{k=1}_{\rm FH}(t)={\rm sin}^{-1}r(t)$,
$\chi^{k=0}_{\rm FH}(t)=r(t)$, and  $\chi^{k=-1}_{\rm FH}(t)={\rm
sinh}^{-1}r(t)$.  For our purpose, we obtain the comoving radial
coordinate $r(t)$,
\begin{equation} \label{3eq3}
r(t)=\frac{1}{\sqrt{|k|}} {\rm sinn}\Bigg[\sqrt{|k|}\chi^{k}_{\rm
FH}(t)\Bigg].
\end{equation}
The definition of $L_{\rm \Lambda}=ar(t)$~\footnote{Definitely,
$L_{\rm \Lambda}=a\chi^{k}_{\rm FH}(t)$ is the proper distance,
while $L_{\rm \Lambda}=ar(t)$ is the radius of the event horizon
measured  on the surface of the horizon to define the proper
surface area~\cite{KT,KL}. In this work, we choose $L_{\rm
\Lambda}=ar(t)$ to define the IR cutoff for non-flat universe.} is
useful for non-flat universe~\cite{HL}, which leads to
\begin{equation} \label{3eq4}
\dot{L}_{\rm \Lambda}=H L_{\rm \Lambda}+a
\dot{r}=\frac{c}{\sqrt{\Omega_{\rm \Lambda}}}-{\rm cosn}y,
\end{equation}
where ${\rm cosn}y=\cos y$ for $k=1$, $y$ for $k=0$, and $\cosh y$
for $k=-1$ with $y=\sqrt{|k|} R_{\rm FH} /a$. Using Eq.
(\ref{3eq3}) together with  $L_{\rm \Lambda}=ar(t)$, we rewrite it
as  ${\rm cosn}y = \sqrt{1 - c^2 \frac{\Omega_{\rm k}}{\Omega_{\rm
\Lambda}}}$ in terms of $\Omega_{\rm k}$ and $\Omega_{\rm
\Lambda}$~\cite{SZZ}.

 Using the definition of $\rho_{\rm
\Lambda}$ and (\ref{3eq4}), one finds the  equation of state
\begin{equation} \label{3eq5}
\dot{\rho}_{\rm \Lambda} +3H \Big[1-
\frac{1}{3}-\frac{2\sqrt{\Omega_{\rm \Lambda}}}{3c}{\rm
cosn}y\Big]\rho_{\rm \Lambda}=0.
\end{equation}
  From Eqs.(\ref{2eq4}), (\ref{2eq6}) and (\ref{3eq5}), we find  the effective equation of state
\begin{equation}
\label{3eq6} \omega^{\rm eff}_{\rm \Lambda}(x)=
-\frac{1}{3}-\frac{2\sqrt{\Omega_{\rm \Lambda}(x)}}{3c}{\rm cosn}y.
\end{equation}  On the
other hand, the effective equation of state for CDM is given
differently by
\begin{equation} \label{3eq7}
\omega^{\rm eff}_{\rm m}(x)=-\frac{b^2}{\Omega_{\rm
\Lambda}^{n-1}}\frac{(1+\Omega_{\rm k})^n}{(1-\Omega_{\rm
\Lambda}+\Omega_{\rm k})}.
\end{equation}
 Now we are in a position to derive two coupled equations whose solutions determine
 the effective equations of state.
Eq.(\ref{2eq3}) leads to one differential equation for
$\Omega_{\rm \Lambda}$
\begin{equation} \label{3eq8}
\frac{d \Omega_{\rm \Lambda}}{dx}=-3\Omega_{\rm
\Lambda}(1-\Omega_{\rm \Lambda}+\Omega_{\rm k})\Big(\omega^{\rm
eff}_{\rm \Lambda}-\omega^{\rm eff}_{\rm m}\Big)+\Omega_{\rm
k}\Omega_{\rm \Lambda}(1+3\omega^{\rm eff}_{\rm \Lambda}\Big)
\end{equation}
with $x=\ln a$. The other differential equation for $\Omega_{\rm
k}$ comes from the derivative of $r_{\rm k}$ in Eq.(\ref{2eq11})
using Eq.(\ref{2eq10}) as
\begin{equation} \label{omegak}
\frac{d \Omega_{\rm k}}{dx}=-3\Omega_{\rm k}(1-\Omega_{\rm
\Lambda}+\Omega_{\rm k})\Big(\omega^{\rm eff}_{\rm
\Lambda}-\omega^{\rm eff}_{\rm m}\Big)+\Omega_{\rm
k}\Big(1+\Omega_{\rm k}\Big)\Big(1+3\omega^{\rm eff}_{\rm
\Lambda}\Big).
\end{equation}
\begin{figure}[t!]
   \centering
\scalebox{.9}
   {\includegraphics{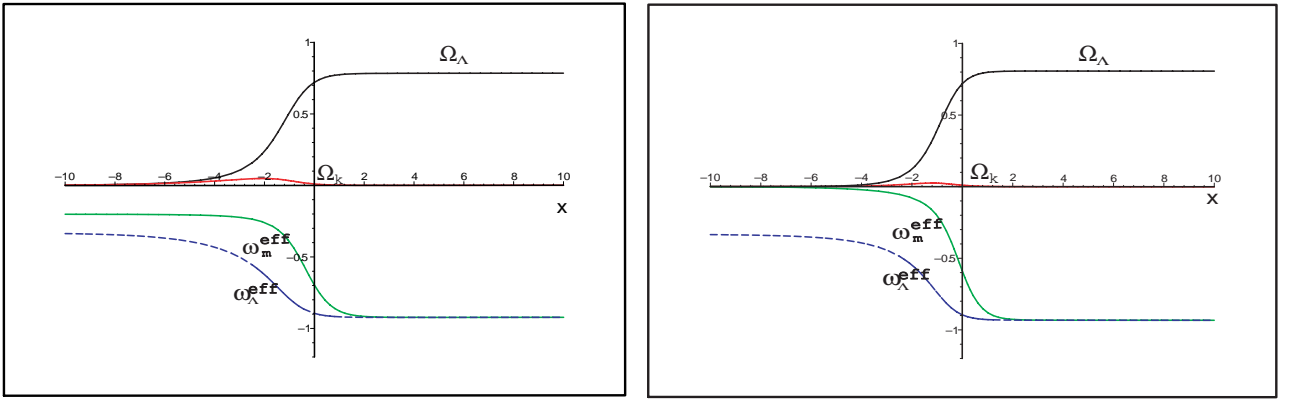}}
\caption{(color online) For $b^2=0.2$ and $c=1$, $k=1$ evolution
of $\Omega_{\rm \Lambda}$ (black) and $\Omega_{\rm k}$ (red) and
the effective equations of state, $\omega^{\rm eff}_{\rm m}$
(green) and $\omega^{\rm eff}_{\rm \Lambda}$ (blue). Here $x=\ln
a$ moves backward ($-$) or forward ($+$) with the present time
$x=0(a_0=1)$. The left picture is for an interaction
 of $n=1$ and the right picture is for $n=1/2$. } \label{fig1}
\end{figure}
At this point, we compare our equations of (\ref{2eq8}),
(\ref{3eq8}) and (\ref{omegak}) with those in~\cite{WLA}. Using
$d\Omega_{\rm k}/dx=\Omega_{\rm k}(1+\Omega_{\rm k})+3\Omega_{\rm
k}\Omega_{\rm \Lambda}\omega_{\Lambda}$, these correspond to (5),
(6), and (8) in~\cite{WLA}, respectively. Hence our model is the
same as in \cite{WLA}. In the case of $\Omega_{\rm k}=0$, equation
(\ref{3eq5}) leads to the well-known form in~\cite{BS}. From Eqs.
(\ref{3eq5}) and (\ref{omegak}), we find a future fixed point of
$\Omega_{\rm \Lambda}$ which satisfies $\omega^{\rm eff}_{\rm
\Lambda}=\omega^{\rm eff}_{\rm m}$ and $\Omega_{\rm k}=0$. If one
drops off the interaction ($\Gamma=b^2=0$), the dark energy
evolution for a flat universe of $k=0$  usually proceeds from the
past fixed point $\Omega_{\rm \Lambda}=0$ to the other future
point $\Omega_{\rm \Lambda}=1$. If the interaction is turned on,
$\Omega_{\rm \Lambda}$ approaches a fixed asymptotic value less
than 1  for large time.
\begin{figure}[t!]
   \centering
 \scalebox{.9}
   {\includegraphics{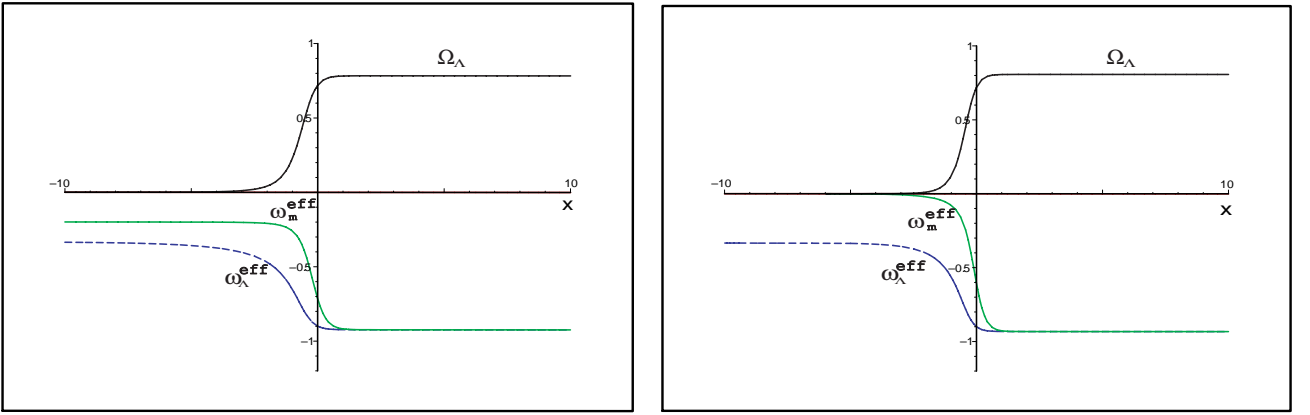}}
\caption{(color online) For $b^2=0.2$ and $c=1$, $k=0$ evolution
of $\Omega_{\rm \Lambda}$ (black)  and the effective equations of
state, $\omega^{\rm eff}_{\rm m}$ (green) and $\omega^{\rm
eff}_{\rm \Lambda}$ (blue). The left picture is for an interaction
 of $n=1$ and the right picture is for $n=1/2$. } \label{fig2}
\end{figure}

In order to obtain solution, we have to solve the above coupled
equations numerically by considering the initial condition at
present time\footnote{Here we use the data from the combination of
WMAP3 plus the HST key project constraint on $H_0$~\cite{Wmap3}.}:
$\frac{d \Omega_{\rm \Lambda}}{dx}|_{x=0}>0,~\Omega^{0}_{\rm
\Lambda}=0.72, \Omega^{0}_{\rm k=1}=0.01/\Omega^{0}_{\rm
k=0}=0.0/\Omega^{0}_{\rm k=-1}=-0.01$.

\begin{figure}[t!]
   \centering
\scalebox{.9}
   {\includegraphics{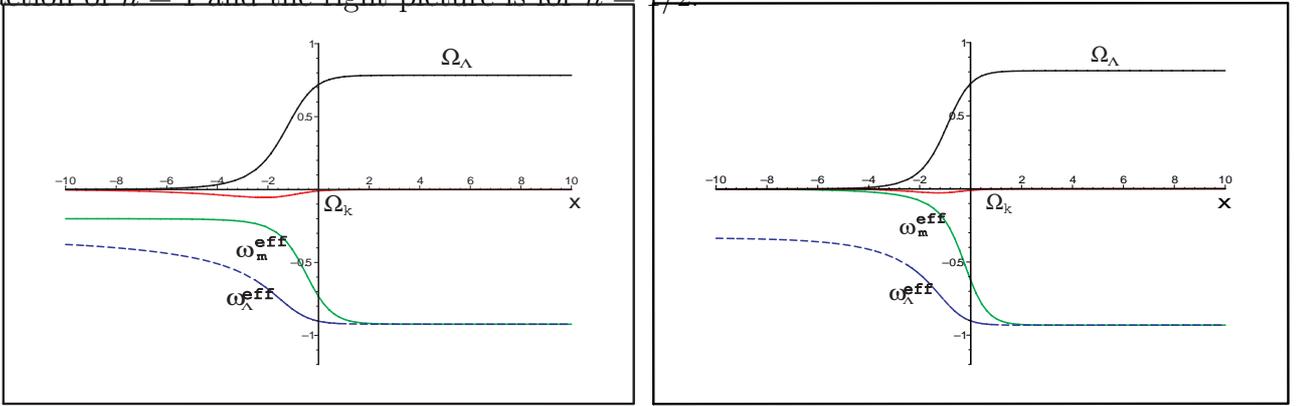}}
\caption{(color online)  For $b^2=0.2$ and $c=1$, $k=-1$ evolution
of $\Omega_{\rm \Lambda}$ (black) and $\Omega_{\rm k}$ (red) and
the effective equations of state, $\omega^{\rm eff}_{\rm m}$
(green) and $\omega^{\rm eff}_{\rm \Lambda}$ (blue). The left
picture is for an interaction
 of $n=1$ and the right picture is for $n=1/2$. } \label{fig3}
\end{figure}

\section{Discussions}

The noninteracting  picture  with $L_{\rm \Lambda}=R_{\rm FH}$ has
the natural tendency such that a ratio $r_{\rm m}$ of two
densities $\rho_{\rm m}$ and $\rho_{\rm \Lambda}$ decreases as the
universe evolves~\cite{LI}. In this case the energy-momentum
conservation is required for each matter separately. Also the
natural tendency holds even for the case including an interaction
between the holographic dark energy  and CDM~\cite{WGA}. They used
the native EoS $\omega_{\rm \Lambda}$ to show that $\rho_{\rm
\Lambda}$  can describe the phantom regime. However, we have to
use the effective EoS $\omega^{\rm eff}_{\rm \Lambda}$ in the
presence of  the interaction. As are shown in Figs. 1, 2, and 3,
two effective equations of state start differently. However, two
effective equations of state will take the same negative value
which is greater than $-1$ in the far future. Also this value
could be estimated from the future fixed point.

\begin{figure}[t!]
   \centering
\scalebox{.9}
   {\includegraphics{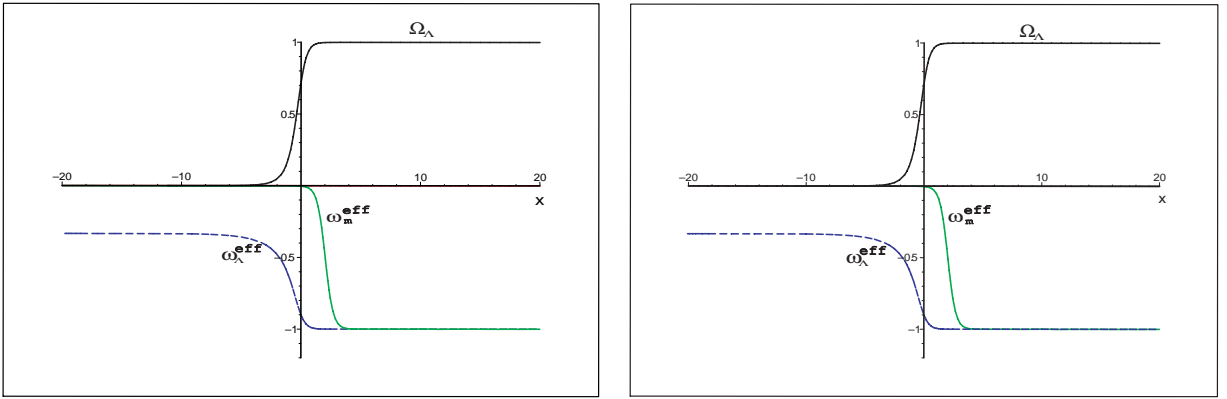}}
\caption{(color online)  For $b^2=0.001$ and $c=1$, $k=0$
evolution of $\Omega_{\rm \Lambda}$ (black) and the effective
equations of state, $\omega^{\rm eff}_{\rm m}$ (green) and
$\omega^{\rm eff}_{\rm \Lambda}$ (blue). The left picture is for
an interaction
 of $n=1$ and the right picture is for $n=1/2$. } \label{fig4}
\end{figure}

The  vacuum decay picture is still alive even for a dynamical
evolution in the interacting holographic dark energy model.  This
implies that one  cannot  generate a phantom-like mixture  of
$\omega^{\rm eff}_{\rm \Lambda}<-1$ from an interaction between
the holographic dark energy and CDM. In other words, decaying from
the holographic dark  energy  into  the CDM never leads to the
phantom regime.  Figs. 1, 2, and 3 show clearly that the density
parameter $\Omega_{\rm \Lambda}$ approaches  0.78 with $b^2=0.2$
and $c=1$, irrespective of the curvature constant $k$ and the
interaction $n$. Furthermore, at $\Omega_{\rm \Lambda}=0$, one
recognizes the changes from $\omega_{\rm m}^{\rm eff}=-0.2$ for
$n=1$ to $\omega_{\rm m}^{\rm eff}=0$ for $n=1/2$. This implies
that a decay rate of $\Gamma=3b^2\sqrt{1+r_{\rm m}}H$ leads to the
CDM-dominated universe with $\omega_{\rm m}^{\rm eff}=0$ at the
far past. We note that the effect of non-flat universe is trivial
because $\Omega_{\rm k}$ goes to zero for the far past and far
future. This means that the non-flat universe of $k\not=0$ could
not induce the phantom phase even one includes an interaction
between the holographic dark  energy and CDM.

We comment on  the fine-tuning and coincidence problems. The
holographic energy density $ \rho_{\rm \Lambda}$ could resolve the
fine-tuning problem because taking $L_{\rm \Lambda}=l_p=1/M_p$
leads to the cosmological vacuum energy $\rho^p_{\rm
\Lambda}\propto M_p^4$. This means that a small system at planck
scale provides an upper limit of $\rho_{\rm \Lambda}\le
\rho^p_{\rm \Lambda}$, as is naively expected in quantum field
theory. On the other hand, as the universe evolves, a larger
system   will have a smaller energy density. This is  a
consequence of the holography. Thus the holographic principle may
reconcile the quantum field theory at planck scale with the
smallness of the present cosmological vacuum energy density
$\rho^0_{\rm \Lambda} \propto M^2_pH^2_0=10^{-123}\rho^p_{\rm
\Lambda}$.

Furthermore, the resulting equilibrium between holographic dark
energy and CDM offers a possible resolution to the cosmic
coincidence problem. The cosmic coincidence problem states that it
is unlikely that the current epoch with  sizable amounts of both
CDM and dark energy coincides with the rapid transition from
CDM-domination to dark energy-domination. Any interacting
holographic models using the future event horizon show the
decreasing effective equations of state. Considering a decay of
the holographic dark energy into CDM, we expect to show the
changes for $k=0$ and $n=1$: $\Omega_{\rm
\Lambda}=0.0~(\Omega_{\rm m}=1.0)$ at the far past; $\Omega_{\rm
\Lambda}=0.72~(\Omega_{\rm m}=0.28)$ at present; $\Omega_{\rm
\Lambda}=0.78~(\Omega_{\rm m}=0.22)$ at the far future. If there
is no interaction, one finds the natural tendency for dark energy
to dominate over CDM as the universe expands: $\Omega_{\rm
\Lambda}=0.0~(\Omega_{\rm m}=1.0)$ at the far past; $\Omega_{\rm
\Lambda}=0.72~(\Omega_{\rm m}=0.28)$ at present; $\Omega_{\rm
\Lambda}=1.0~(\Omega_{\rm m}=0.0)$ at the far future. This means
that the interaction makes  the slow transition from
CDM-domination to dark energy-domination. The natural tendency is
compensated by the decay of the holographic dark energy into
CDM~\cite{BS}. As is expected from the future fixed point of
$\omega_{\rm \Lambda}^{\rm eff}=\omega_{\rm m}^{\rm eff}$ , there
exist a balance between tendency and decay. Thus we have the
effective equation of state $\omega_{\rm \Lambda}^{\rm
eff}=-0.92(n=1)$ and $\omega_{\rm \Lambda}^{\rm eff}=-0.93(n=1/2)$
for an equilibrium mixture.

At this stage, we consider  a very weakly coupling of $b^2$.
According to the interacting quintessence models~\cite{Pavon},
$b^2$ corresponds to the parameter $c_{\rm OAP}^2$ which must be
lower than 0.001. In this case, if $c_{\rm OAP}^2 > 0.001$, a
baryon-dominated universe would develop before the dark
matter-domination. This would hinder tremendously the formation of
cosmic structure. In order to see whether this picture is possible
to occur within the interacting holographic model, we choose
$b^2=0.001$. For $k=0$ and $c=1$ case, we observe its evolution
from Fig. 4. It shows the nearly same form as in Fig.2 except that
$\omega_{\rm \Lambda}^{\rm eff}=-0.98(n=1)$ and $\omega_{\rm
\Lambda}^{\rm eff}=-0.98(n=1/2)$ for an equilibrium mixture. This
means that the nature of holographic interaction is not changed
even for a very small coupling of $b^2$. Therefore, we could not
find  such a condition of $c_{\rm OAP}^2$ in our model. The only
limitation on $b^2$ comes from the condition of the natural
tendency for dark energy: $d\Omega_{\rm \Lambda}/dx|_{x=0}>0 \to
b^2<b^2_{\rm max}$, where $b^2_{\rm max}$ satisfies $d\Omega_{\rm
\Lambda}/dx|_{x=0}=0$. As an example, we have $b^2_{\rm max}=0.35$
for $k=0,c=1,n=1,$ and $ \Omega^0_{\rm \Lambda}=0.72$.

In addition we have three parameters $b^2, c, n$ and observational
ranges on $\Omega_{\rm \Lambda}, \Omega_{\rm m},\Omega_{\rm k}$.
Hence it suggests  that there is a parameter space which may
describe a phantom-regime of $\omega_{\rm \Lambda}^{\rm eff}<-1$.
However, it is easily proved  that this is not the case. Requiring
the second-law of thermodynamics of $\dot{S}_{BH}=2 \pi R_{\rm FH}
\dot{R}_{\rm FH} \ge 0$ leads to the condition of $c \ge
\sqrt{\Omega_{\rm \Lambda}}{\rm cosn}y$. On the other hand,  the
condition for $\omega_{\rm \Lambda}^{\rm eff}<-1$ with
Eq.(\ref{3eq6}) implies that $c < \sqrt{\Omega_{\rm \Lambda}}{\rm
cosn}y$. Hence two conditions are not compatible. An important
parameter to determine $\omega_{\rm \Lambda}^{\rm eff}$ is $c$.
Actually the interaction ($b^2$ and $n$) between holographic dark
energy and CDM  dose not induce a phantom-like matter.

 Finally, we mention the recent observations. A lot of data
 support on the flat universe. Also it would be important to
 stress on the motivation of considering the non-flat universe
 with the small $\Omega_{\rm k}$ from CMB experiments~\cite{Wmap1,Wmap3} and
 supernova measurements~\cite{SN}. Our results show that  the
 effect of the non-flat universe becomes trivial because $\Omega_{\rm
 k}$ goes to zero for the far past and the far future, even the non-flat
 universe contributes small at present. Hence,  if the interacting holographic dark energy model is reliable,
 we anticipate  that the curvature term $\Omega_{\rm k}$  does not play an important role
 for determining the future dark energy-dominated  universe.

 Consequently, it turned out
that the interacting holographic energy density could not describe
the phantom regime\footnote{However, it seems that there were
alternatives to provide the phantom phase. The general
interaction~\cite{SH}, the particle horizon ~\cite{Sim}, and $c<1$
case~\cite{SZZ} were used to derive phantom phase.}.

\section*{Acknowledgment}
 K. Kim and H. Lee were  in part supported by
KOSEF, Astrophysical Research Center for the Structure and
Evolution of the Cosmos at Sejong University. Y. Myung  was in
part supported by the SRC Program of the KOSEF through the Center
for Quantum Spacetime (CQUeST) of Sogang University with grant
number R11-2005-021 and by the Korea Research Foundation
(KRF-2006-311-C00249) funded by the Korea Government (MOEHRD).

\end{document}